\documentclass[10pt,conference]{IEEEtran}

\IEEEoverridecommandlockouts
\usepackage{cite}
\usepackage{amsmath,amssymb,amsfonts}
\usepackage{algorithmic}
\usepackage{enumitem}
\usepackage{graphicx}
\usepackage{textcomp}
\usepackage{url}
\usepackage{array}
\usepackage{xcolor}
\usepackage{lipsum}
\usepackage{svg}
\usepackage{titlesec}
\usepackage{hyperref}
\usepackage{float}
\usepackage{booktabs}
\usepackage{tabularx}
\usepackage{float}
\usepackage{tcolorbox} %
\usepackage{balance}
\usepackage{listings} %

\begin{document}

\title{
An Empirical Study on Bugs Inside PyTorch: A Replication Study
}

\author{
\IEEEauthorblockN{Sharon Chee Yin Ho$^*$}
\IEEEauthorblockA{
    \textit{Concordia University}\\
    sharoncheeyin.ho@concordia.ca
}
\and
\IEEEauthorblockN{Vahid Majdinasab$^*$}
\IEEEauthorblockA{
    \textit{Polytechnique Montréal}\\
    vahid.majdinasab@polymtl.ca
}
\and
\IEEEauthorblockN{Mohayeminul Islam$^*$}
\IEEEauthorblockA{
    \textit{University of Alberta}\\
    mohayemin@ualberta.ca
}
\and
\IEEEauthorblockN{Diego Elias Costa}
\IEEEauthorblockA{
    \textit{Université du Québec à Montréal}\\
    costa.diego@uqam.ca
}
\and
\IEEEauthorblockN{Emad Shihab}
\IEEEauthorblockA{
    \textit{Concordia University}\\
    emad.shihab@concordia.ca
}
\and
\IEEEauthorblockN{Foutse Khomh}
\IEEEauthorblockA{
    \textit{Polytechnique Montréal}\\
    foutse.khomh@polymtl.ca
}
\and
\IEEEauthorblockN{Sarah Nadi}
\IEEEauthorblockA{
    \textit{University of Alberta}\\
    nadi@ualberta.ca
}
\and
\IEEEauthorblockN{Muhammad Raza}
\IEEEauthorblockA{
    \textit{Queen's University}\\
    m.raza@queensu.ca
}}

\maketitle
\def\thefootnote{*}\footnotetext{These authors contributed equally to this work}

\thispagestyle{plain}
\pagestyle{plain}

\begin{abstract}
Software systems are increasingly relying on deep learning components, due to their remarkable capability of identifying complex data patterns and powering intelligent behaviour.
A core enabler of this change in software development is the availability of easy-to-use deep learning libraries.
Libraries like PyTorch and TensorFlow empower a large variety of intelligent systems, offering a multitude of algorithms and configuration options, applicable to numerous domains of systems. 
However, bugs in those popular deep learning libraries also may have dire consequences for the quality of systems they enable; thus, it is important to understand how bugs are identified and fixed in those libraries.  

Inspired by a study of Jia et al., which investigates the bug identification and fixing process at TensorFlow, we characterize bugs in the PyTorch library, a very popular deep learning framework.
We investigate the causes and symptoms of bugs identified during PyTorch's development, and assess their locality within the project, and extract patterns of bug fixes.
Our results highlight that PyTorch bugs are more like traditional software projects bugs, than related to deep learning characteristics.
Finally, we also compare our results with the study on TensorFlow, highlighting similarities and differences across the bug identification and fixing process.

\end{abstract}


\section{Introduction}

The recent advances in deep learning research allied to the wide availability of large datasets have pushed deep learning components into our everyday systems.
Deep learning-powered systems have successfully been deployed in multiple domains, from medicine~\cite{Wang:HealthSystems} to self-driving cars~\cite{bojarski2016end}, to automate highly complex tasks due to their capabilities of learning complex data patterns. 
A major catalyst for this widespread use of deep learning components is the availability of mature and easy to use deep learning libraries. 
Libraries like PyTorch~\cite{paszke2019pytorch} and TensorFlow~\cite{abadi2016tensorflow} abstract much of the complexity of defining, training, and testing deep learning components with a high-level functional API, and are becoming widely popular.
In the Stack Overflow survey of 2022, deep learning libraries like PyTorch were reportedly used by 11\% of all participants of the survey~\cite{StackOve55:online}.

While very popular DL libraries help empower (artificially) intelligent behaviour in a myriad of systems, bugs in those libraries may have dire consequences. 
Defects in the underlying DL library could eventually propagate as observable failures in the systems that depend on it. 
In fact, many authors in prior studies done on bugs in software systems using DL techniques, \cite{zhang2018empirical,islam2019comprehensive,humbatova2020taxonomy} suggest that such bugs may be impacted by defects found in the DL libraries used as studied by Tambon et al. \cite{tambon2021silent}. %
Therefore, understanding and identifying defects in DL libraries are essential for ensuring the quality of the systems that rely on them.

Two studies from Jia et al.~\cite{jia2021symptoms,jia2020empirical} pioneered an investigation on bugs inside the TensorFlow library. 
The authors provided a comprehensive understanding of potential links between \textit{symptoms} %
and \textit{root causes} of bugs inside TensorFlow, in addition to analyzing which \textit{repair patterns} were used when fixing bugs. To further build and strengthen our empirical knowledge in the area, we define symptoms as the faulty behaviour of a software that indicates a possibility of a bug in the program, such as program crashes or a functionality not providing expected results, with root causes being the reason behind why such a bug has appeared, such as the wrong implementation of an algorithm or a mis-configuration of the software. In addition, we define repair patterns as reoccurring patterns used to fix a bug, such as adding a value checker or throwing an exception when applicable.

In this paper, we replicate the study conducted by Jia et al.~\cite{jia2021symptoms} on the PyTorch library \cite{paszke2019pytorch}, another popular Python library for developing deep learning systems. PyTorch is used in many critical applications, with Tesla's Autopilot \cite{pytorchtesla} and Uber's Pyro \cite{goodman2019uber} being notable examples. We argue that this replication study on PyTorch which complements the work done by Jia et al. on TensorFlow, will benefit the machine learning community by having a comprehensive overview of the bug characteristics and its corresponding repair patterns inside the two most popular DL libraries. In addition, by having three similar research questions as Jia et al. \cite{jia2021symptoms}, we can have a direct comparison to see the similarities and differences of bugs in terms of the symptoms, causes, components affected, and repair patterns. We find that while some results in PyTorch are similar to those found in TensorFlow \cite{jia2021symptoms}, there are also dissimilarities. \textbf{For example, we find that \textit{crash} and \textit{functional error} are most frequent symptoms of bugs, which matches their findings. However, we observe \textit{processing} as a low occurring root cause, which is the top root cause in the TensorFlow study.}

Our contributions in this paper are as follows:
\begin{itemize}
    \item We study 194 bugs in the popular DL library PyTorch to understand the bug characteristics and how the bugs are fixed.
    \item We compare our results with a similar study done on TensorFlow to outline the similarities and differences between these two frameworks.
    \item To promote further research in the area, we also share our replication package, containing all bugs analyzed, our coding scheme, and more detailed results~\footnote{\url{https://github.com/datasetsharing/pytorchbugdataset}}.
\end{itemize}
The rest of this paper is structured as follows. In the next Section, we discuss the related works. In Section \ref{study_design} we outline our study's design, explain our research questions and our data collection and curation method. In Sections \ref{rq1}, \ref{rq2}, and \ref{rq3} we describe our findings for each of our research questions. In Section \ref{discussion}, we discussed the implications of our study. In Section \ref{threats_to_validity}, we outline the threats to our results' validity, and finally in Section \ref{conclusion} we conclude our paper and propose areas for future work.

\section{Related Works}\label{related_works}
In this section, we outline the current work that has been done on defect analysis in machine learning and deep learning systems. 
    
    \subsection{Understanding Defects in Traditional Systems}
    Pradel et al. \cite{pradel2018deepbugs} designed a DL approach for bug detection by analyzing variable names. Lu et al. \cite{lu2005bugbench} created a dataset of bugs for benchmarking bug detection tools.  Wang et al. \cite{wang2016bugram} proposed an approach for bug detection by using n-gram language models instead of rule based approaches. Pan et al. \cite{pan2006bug} introduced program slicing metrics for measuring coupling, cohesion, and complexity of codes written in C language. %
    
    Zhu et al. \cite{zhou2016combining} proposed combining techniques from text mining and data mining for bug prediction. D'ambros et al. \cite{d2010extensive} have designed another bug benchmarking dataset for evaluating bug prediction tools. Hammouri et al. \cite{hammouri2018software} used three supervised learning approaches (Naive Bayes, Decision Trees, and Neural Networks) for bug prediction. Arcuri et al. \cite{arcuri2008novel} proposed an approach based on co-evolution where the program and test cases evolve alongside each other for bug fixing. Ye et al. \cite{ye2021comprehensive} did a comprehensive study on automatic repair tools on the QuixBugs benchmark.

    \subsection{Understanding Defects in Machine Learning Systems}
    Thung et al. \cite{thung2012empirical} analyzed the bug database of three ML-based systems and labeled the bugs into different categories. They also analyzed the relationships between different dimensions of bugs, such as the category, severity, cost of fixing, impact, etc. Kim et al. \cite{kim2021denchmark} developed a benchmark of bugs in ML systems with a focus to help with automatic debugging. In addition, some empirical studies have also been done on bugs in applications of a branch of ML known as deep learning.
    
    Y. Zhang et al. \cite{zhang2018empirical} conducted an empirical study of bugs in applications that call the APIs of TensorFlow \cite{abadi2016tensorflow} and their corresponding fixes to find the root causes of the bugs. Islam et al. \cite{islam2019comprehensive} analyzed applications using DL libraries, such as Caffe \cite{jia2014caffe}, Keras \cite{keras}, Theano \cite{bergstra2011theano}, and Torch \cite{collobert2002torch}. Humbatova et al. \cite{humbatova2020taxonomy} also conducted a similar study on applications using TensorFlow \cite{abadi2016tensorflow}, Keras \cite{keras} and PyTorch \cite{paszke2019pytorch}. X. Zhang et al. \cite{zhang2020towards} studied the bugs related to incorrect decisions provided by DL systems and characterized such bugs.
    Finally, J. Chen et al. \cite{chen2022toward} conducted a study of bugs on four DL frameworks (TensorFlow, PyTorch, MXNET, and DL4) by analysing 1000 bugs. They analyze the root causes and symptoms of these bugs alongside the components in which they appeared in.
    
    By analyzing the related works mentioned above, we observe that there have been many studies done on analyzing defects in traditional software systems, ML/DL systems, and software which incorporate ML/DL components. However, there has been very little study on the DL frameworks themselves. 
    To the best of our knowledge, there are only two works done in this area. An empirical study done by Jia et al. \cite{jia2021symptoms} on bugs inside TensorFlow and a study done by Chen et al. \cite{chen2022toward} on bugs on 4 different DL frameworks. Alongside TensorFlow, PyTorch is also a popular DL framework that is used in research and industry alike. Our aim of doing a replication study on PyTorch based on the study that investigated TensorFlow by Jia et al. \cite{jia2021symptoms}, is to draw direct comparisons between the two most popular DL libraries and to observe the defects that are most prevalent in developing DL systems.
    
    The study done by Chen et al. \cite{chen2022toward} has similarities to our work as they study TensorFlow and PyTorch alongside MXNET and DL4J. However, there exist differences in the depth and scope of analysis of both works. As mentioned, our work is a replication of the study done by Jia et al.\cite{jia2021symptoms} to directly compare the different root causes, symptoms, and bug locations between TensorFlow and PyTorch. We also investigate the different repair patterns that are used to address these bugs by PyTorch's development team, which is not present in the work of Chen et al.~\cite{chen2022toward}. 
    We believe our work complements well the literature cited above and contributes to building an even stronger empirical foundation about bugs on DL framework. 
    By replicating the study of Jia et al.~\cite{jia2021symptoms}, our study provides a direct and detailed comparison of bugs and their fixing process on the two most popular DL frameworks, PyTorch and TensorFlow.

\section{Study Design}\label{study_design}

    Our goal in this study is to better understand the characteristic of bugs in DL libraries and their respective bug-fixing practices. Therefore, we aim to answer the following research questions:
    \begin{itemize}
        \item\textit{RQ1: What are the symptoms and root causes of bugs in PyTorch?}
        \item\textit{RQ2: Which PyTorch components are prone to bugs?}
        \item\textit{RQ3: What are the repair patterns inside PyTorch?}
    \end{itemize}
     We compare our findings with the TensorFlow paper by Jia et al. \cite{jia2021symptoms}. 
     Since both PyTorch and TensorFlow are DL libraries and written in the same languages, there are similar components and structures. %
     To be able to do a fair comparison, we categorize the bugs into similar root causes and symptoms, as well as bugs locations and repair patterns as done by the TensorFlow paper. We then compare their distributions to have a better understanding of their similarities and differences. In addition, when we find some dissimilarities we mention them in the results section.

    \subsection{Data Collection and Curation}
    We aim to analyze bug reports, which can be found as issues inside PyTorch's GitHub repository. To gather these bug reports, we first automatically mine candidate issues for analysis using GitHub's API. Then we sorted the issues by reaction count. After that, we manually filtered out irrelevant issues. Finally, we labelled the issues for further analysis to answer our research questions. Our methodology for collecting the dataset is a bit different from that of Jia et al. \cite{jia2021symptoms}, where instead of searching for bugs that might correspond to pull requests merged into the master branch, we look directly at reported issues that have been fixed via a corresponding pull request merged into the master branch.
    
    Figure \ref{fig:method} shows an overview of our data collection process. We further define and discuss the steps in the subsections below:
        
    \begin{figure*}
        \centering
        \includegraphics[width=0.7\textwidth]{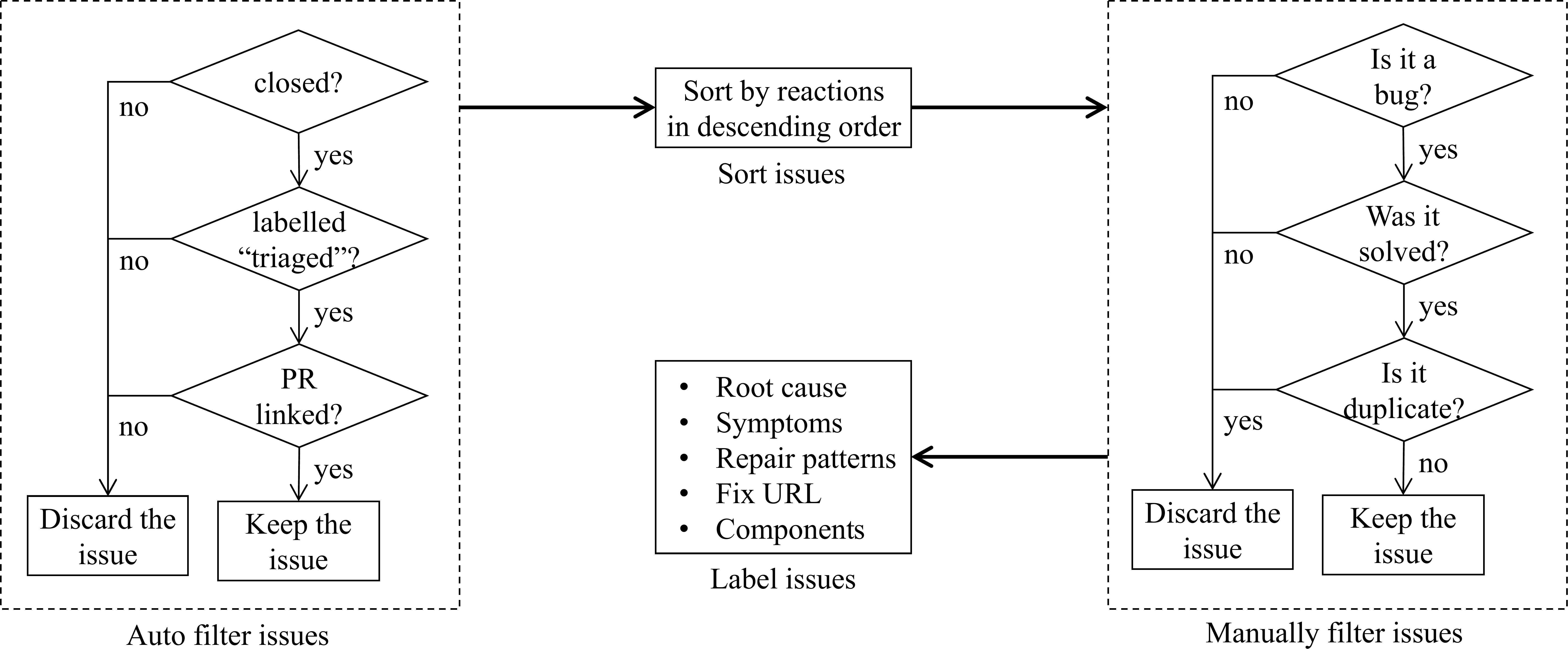}
        \caption[]{Overview of the data collection process of our study.}
        \label{fig:method}
    \end{figure*}

        \subsubsection{Auto filter issues}
        As of 20 October 2022, PyTorch had 19,373 closed issues\footnote{\url{https://github.com/pytorch/pytorch/issues?q=is:issue+is:closed}}, many of which include bug reports.
        Since it is not feasible to manually review all the issues in PyTorch's GitHub issue tracker, we use GitHub API to keep a list of probable bug reports relevant to our study. We used the following criteria for filtering, which yielded a total of 2205 issues:

        \begin{itemize}
            \item\textbf{\textit{Closed issues.}} We are only interested in closed issues because we want to only investigate resolved bugs. We cannot find the root cause and repair patterns for non-closed issues.
            
            \item\textbf{\textit{Labeled as ``triaged''.}} We studied the PyTorch issue labels to understand how they use them. To choose the appropriate label to extract relevant issues for our study, we first sorted the list of labels used in the project by most issues. This yielded ``triaged`` \footnote{\url{https://github.com/pytorch/pytorch/labels?q=triaged}} as the most used label, with its description being: "This issue has been looked at a team member, and triaged and prioritized into an appropriate module". This label was then chosen because this label indicates that the bug report has been reviewed by a project member and has been manually allocated to an appropriate module for further investigation. This also indicates that the issue is relevant to the project and may present an actual bug in the PyTorch repository, which would be notable for our study.
            \item\textbf{\textit{Having a linked pull request.}} If the closed issue has pull requests linked to it, we assume that the pull request was done to fix the bug. The pull requests for a bug are found by looking at the issue itself and seeing that it has been manually linked to existing pull requests that fixes the bug reported in the issue.
        \end{itemize}
        
        \subsubsection{Manually filter the issues}
        After the initial filtering, the first 3 authors sorted the remaining issues for priority review during a manual analysis session, by descending order of the number of reactions. 
        Reactions are a feature in GitHub for developers to respond to a comment or issue using different emojis (i.e., +1, -1, laugh, confused, heart, hooray, rocket, eyes). The number of reactions gives some idea on how important the issue was to developers and community members, which can indicate important bug reports and issue discussions in the project for the community.
        We started by manually filtering
        the issues from the top of the sorted list of issues.
        We discard an issue if it is a duplicate of another issue we already considered.
        Other than bug reports, PyTorch developers use the issue tracker to log feature requests, documentation requirements, and many other types of issues.
        The issue description usually mentions whether or not the issue is a bug report. We discard issues that are not bug reports or were reported by a community member but other developers were not able to replicate it.
        We read through the discussions of each issue and decided whether it was closed because it was fixed or for some other reason.
        We only kept bug reports which were resolved and fixed by a member of PyTorch's development team.
        As we aim to replicate the study done by \cite{jia2021symptoms}, and to have comparable results to their 202 analyzed bugs, we kept the top 200 issues that were left after automatically and manually filtering the issues for analysis.

        \subsubsection{Label the issues}
        After confirming the dataset, we analyzed the 200 selected issues to identify their symptoms, potential root causes (RQ1), and components affected (RQ2). We looked at their corresponding pull requests to identify the repair pattern or actions taken to resolve the bug (RQ3). During this process and upon further analysis, we found 6 issues out of the original 200 to not be bug reports, therefore, we discarded them. Therefore our analysis is on the 194 remaining bug reports.
        
        We look at the following information of the bug reports to label them:
        \begin{itemize}
            \item\textbf{\textit{Issue labels.}} While looking through PyTorch's issue tracker on GitHub, we found that the symptoms, causes, and components affected are sometimes explicitly mentioned in the issue labels. Examples of such labels include \textit{"module: crash"}, \textit{"module: build warnings"}, \textit{"module: deadlock"}, to name a few. These are useful when finding answers to RQ1 and RQ2, as they already provide potential context to the characteristic of the bug itself.
            \item\textbf{\textit{Issue description.}} The first comment in the GitHub issues is considered the issue description. The symptoms, causes, and components affected are sometimes indicated in the description.
            \item\textbf{\textit{Other comments.}} The developers and the users often discuss the issues in the comment thread. These discussions help reveal some useful information on the bug characteristics.
            \item\textbf{\textit{Linked issues.}} Some issues are often related and systematically linked within the issue tracker. We checked the linked issues to find the root causes if they cannot be found within the issue.
            \item\textbf{\textit{Linked pull requests.}} Issues are often linked to pull requests and pull requests have their own discussions. These discussions can help identify the components in which the bug was located and what repair patterns were used to resolve the bug.
            \item\textbf{\textit{Linked commits.}} Issues often have links to the commits that were done to fix the problem, which are found in pull requests. These code changes in such commits help to reveal the bug location and repair patterns.
        \end{itemize}
    
    \subsection{Qualitative Analysis Methodology}\label{methodology_section}
    We outline our methodology in this section. From the collected dataset of 194 bugs, we then investigated their causes, symptoms, components affected, and repair patterns. Finally, we compare our findings with existing studies, especially that of Jia et al. \cite{jia2021symptoms} on TensorFlow.
    
        We decided to use the same coding scheme as Jia et al.~\cite{jia2021symptoms} to classify the bugs' symptoms, root cause, and repair patterns.
        We follow closely the interpretation given for each coding category presented by the replicated study, and apply them to the best of our capabilities to the PyTorch project. 
        The goal is to have comparable results between the results of both libraries, which would give us a more robust empirical evidence of the bugs in deep learning libraries.

        The first three authors and the last author worked on analyzing the bugs in the PyTorch project, and are hereby referred to as annotators.   
        Two annotators of this study have a background in machine learning (ML) while the other two are experienced in software engineering (SE) with each annotator having at least 5 years of experience in their respective field. The annotators all had comparable seniority in their respective fields to avoid biasing the annotations towards the opinion of the most senior analyst.
        Each bug is analyzed by at least two annotators from different backgrounds (one from ML and from SE), to identify the root causes, symptoms, and repair pattern. 
        Given we reuse the coding scheme from Jia et al.~\cite{jia2021symptoms}, and our annotators are purposefully paired based on the complement of expertises, we opted for the method of collaborative coding analysis~\cite{Andrew:CollaborativeCoding}.
        In collaborative coding analysis, annotators work together to classify, discuss divergences and reach a consensus in the coding.   
        In some classifications, however, a consensus could not be reached by the two annotators. 
        In such cases, all authors discussed reaching a unified consensus.

\section{Symptoms and Root Causes of Bugs}\label{rq1}

    \subsection{Motivation}
    When developing a DL library, it is crucial to understand the symptoms and causes of bugs. PyTorch is one of the popular DL libraries used by many real-world applications around the globe. Therefore, if symptoms and causes of bugs are not identified in PyTorch, it could potentially lead to crucial failures in the corresponding applications, causing large-scale damage.
    There are insufficient empirical studies on the bug characteristics of DL libraries, particularly in the case of PyTorch. Our research is the first to answer this question for this DL library in particular. This study adds to the work on TensorFlow by Jia et al. \cite{jia2021symptoms} to provide a bigger picture of the overall bug characteristics of major DL libraries. %
        
    \subsection{Methodology}

    At first, we attempt to extract the symptom and root cause from the title and description from the bug report. 
    Often the symptom and root cause are not understood from just the description, leading us to look in the discussion among the bug reporter, developers and other contributors. 
    We also look at the description and discussion in the corresponding pull requests of these bug reports, to determine the root cause and symptoms. For example, the pull request of \#46983 in Listing \ref{bug_46983} is titled ‘‘Make sure valid ParameterList/Dict don't warn on creation’’. Based on the title, we determined that the symptom of the bug is that a warning was raised.
    
    We followed this conversation until we reached the part of the discussion which contained the commit that closed the issue or contained a fix for it. The developers noticed that the only code modification for this bug fix was developing a function for cleanup.
    By looking at the fix, we can determine what the root cause was and what was the approach (repair pattern) taken to resolve it.
    We manually analyzed 194 bugs in PyTorch to find the symptoms and root causes.
    It should be noted that following Jia et al.'s work with TensorFlow bugs \cite{jia2021symptoms}, each bug has one root cause and one primary symptom.

    \begin{lstlisting}[language=Python,basicstyle=\footnotesize,caption={Solution for the bug 46983.},captionpos=b,label=bug_46983]
- if not isinstance(value, torch.nn.Parameter):
+ if getattr(self, "_initialized", False) and not isinstance(value, torch.nn.Parameter): 
    \end{lstlisting}

    \begin{figure*}
        \centering
        \includegraphics[width=0.9\linewidth]{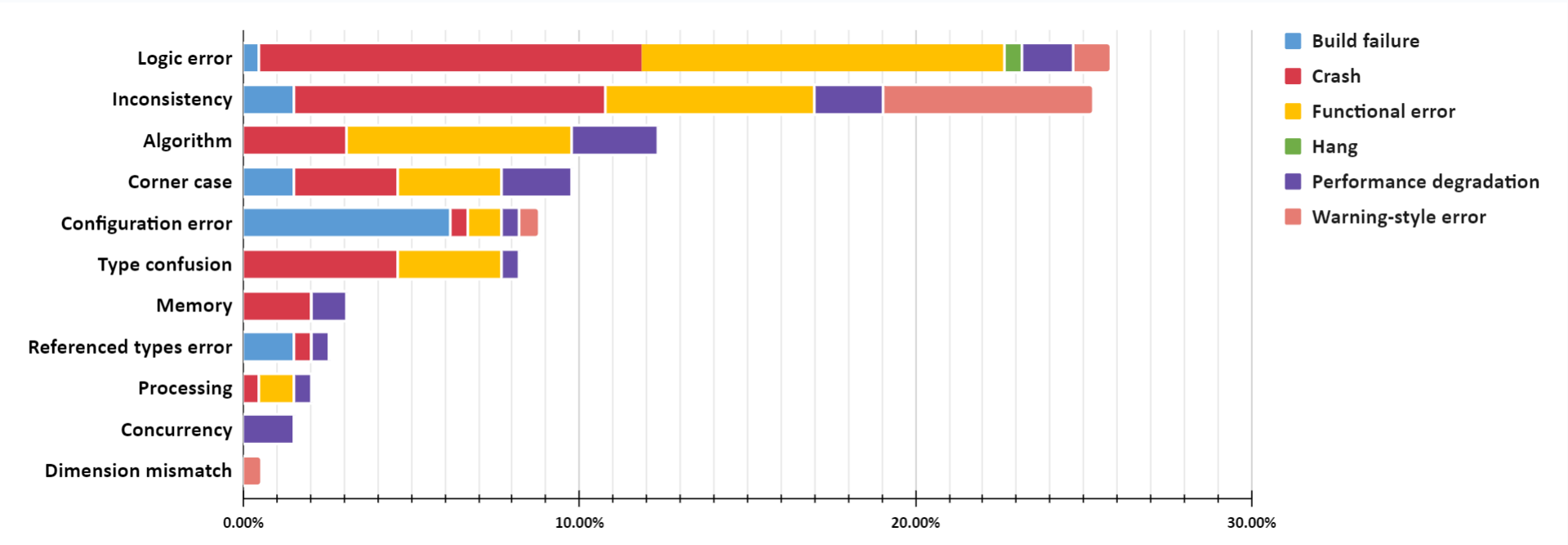}
        \caption{The distribution of symptoms per root cause. }
        \label{fig:distribution}
    \end{figure*}

    \subsection{Results: Root Causes of PyTorch Bugs}
    Following the replicated study on TensorFlow \cite{jia2021symptoms}, we classified the analyzed bugs' root causes into 1 of 11 categories. Our results show that more than 25\% of the bugs analyzed were caused by inconsistencies in the APIs which demonstrate that PyTorch requires more time and development effort in order to be a truly reliable framework. In the following, we discuss the 11 categories for root causes of bugs in PyTorch from the 194 bugs analyzed.

    \begin{table}
        \centering
        \caption{Root causes of bugs identified in PyTorch.} %
        \label{tab:root_causes}
        
\begin{tabularx}{\linewidth}{l X r}
\toprule
   \textbf{Root Cause} & \textbf{Description} & \textbf{Freq.} \\
\midrule
    Logic Error & Wrong programming logic & 25.77\%  \\
    Inconsistency  & Inconsistent changes in the API & 25.26\%  \\
    Algorithm & Wrong implementation of algorithms & 12.37\% \\
    Corner case & Wrong handling of corner cases & 9.79\% \\
    Configuration error & Wrong configurations & 8.76\% \\
    Type confusion & Type mismatches & 8.25\% \\
    Memory & Incorrect usage of memory & 3.09\% \\
    Referenced type error & Incorrect import of libraries & 2.58\% \\
    Processing & Incorrect variable initialization or assignment & 2.06\% \\
    Concurrency & Synchronization problems & 1.55\% \\
    Dimension mismatch & Dimension mismatch between tensors & 0.52\% \\
\bottomrule
\end{tabularx}

    \end{table}
    
    \begin{enumerate}
        \item\textbf{Logic error (25.77\%).} The bugs in this category were caused by wrong programming logic. For example, in issue \#50663~\cite{issue3307:copydeep21}, maintainers report a bug in the implementation of a deep copy operation. A deep copy operation is expected to create an exact replica of the copied object, however, a wrong logic in the implementation caused it to not copy part of the object (gradient buffer), causing users to experience undefined behavior errors.
        
        \item\textbf{Inconsistency (25.26\%).} The bugs in this category were caused by changing the APIs or updating the framework's version which resulted in inconsistencies or incompatibilities between framework interfaces, modules, or functions. For example, pull request (PR) \#53424 \cite{pr53424} reports a bug in calling a tensor object. This bug was caused because of name shadowing after adding a new module in an update which raised an error during creating new tensor objects.
        
        \item\textbf{Algorithm (12.37\%).} The bugs in this category were caused by wrong implementation of algorithms, such as incorrect algorithmic logic or incorrect implementation of algorithmic concepts. For example, in PR \#56488 \cite{pr56488} NumPy was reported to generate the same random number for each data batch as opposed to the expected behaviour. NumPy is actively used in PyTorch's implementation and an incorrect usage of its API had caused this bug.
        
        \item\textbf{Corner case (9.79\%).} The bugs in this category, were caused by wrong handling of corner cases. Corner cases are considered particular use-cases or program execution flow that are not generally used or triggered by library users, but must, nevertheless, be handled by the library. For example, in issue \#16532 \cite{issue16532}, it was reported that gradients are missing when \texttt{autograd} is called inside a function on Multi-GPUs. We classify such issues as corner cases since most developers will not use PyTorch functions in such a way.
        
        \item\textbf{Configuration error (8.76\%).} The bugs in this category were caused by wrong configurations. For example, issue \#22389 \cite{issue22389} reports a bug which caused the developers to be unable to use TensorBoard. This bug happened because a dependency which was required for TensorBoard's functionality was not installed during PyTorch installation.
        
        \item\textbf{Type confusion (8.25\%).} The bugs in this category were caused by type mismatches. Such issues present errors that stop the program from functioning. For example, issue \#42218 \cite{issue42218} reports that the program failed to function because of such an error.
        
        \item\textbf{Memory (3.09\%).} The bugs in this category were caused by incorrect usage of memory resources. These issues can be caused because of using too much RAM or memory leaks. For example, issue \#35901 \cite{issue35901} reports that program failed during run because of an out of memory error.
    \end{enumerate}
     
    The rest of the root cause categories were not prevalent enough to be a cause of concern. Referenced type errors (2.58\%) such as issue \#42435 \cite{issue42435} were caused by incorrect \texttt{include} or \texttt{import} statements. Processing type (2.06\%) errors such as issue \#49052 \cite{issue49052} were caused by program variables being initialized or assigned incorrectly, using incorrect formats for variables, or other incorrect data processing related usages. Concurrency (1.55\%) and dimension mismatch (0.52\%) type errors were caused by synchronization problems (such as issue \#67626 \cite{issue67626}) and dimension mismatch during tensor computation and transformation operations (such as PR \#71065 \cite{issue71065}), respectively.

    Figure \ref{fig:distribution} shows which symptoms a root cause can have based on our findings, providing a detailed view of which symptoms and root cause relationship. 
    Most notable from our results is that Hang is only present in bugs stemming from Logic errors, albeit making up only a small proportion of its symptoms. Logic error bugs, on the other hand, manifest through all symptom types found in this study. Functional error and Crash are the most frequent symptoms for Logic error bugs, making up altogether over half of the total symptoms for Logic error bugs. Another interesting finding is that Concurrency issues only display Performance degradation as a symptom. In addition, the only symptom of Dimension mismatch in our study were Warning-style errors and configuration error bugs also most often display as Build failures.

    \begin{figure}[htb]
        \centering
        \includegraphics[width=\linewidth]{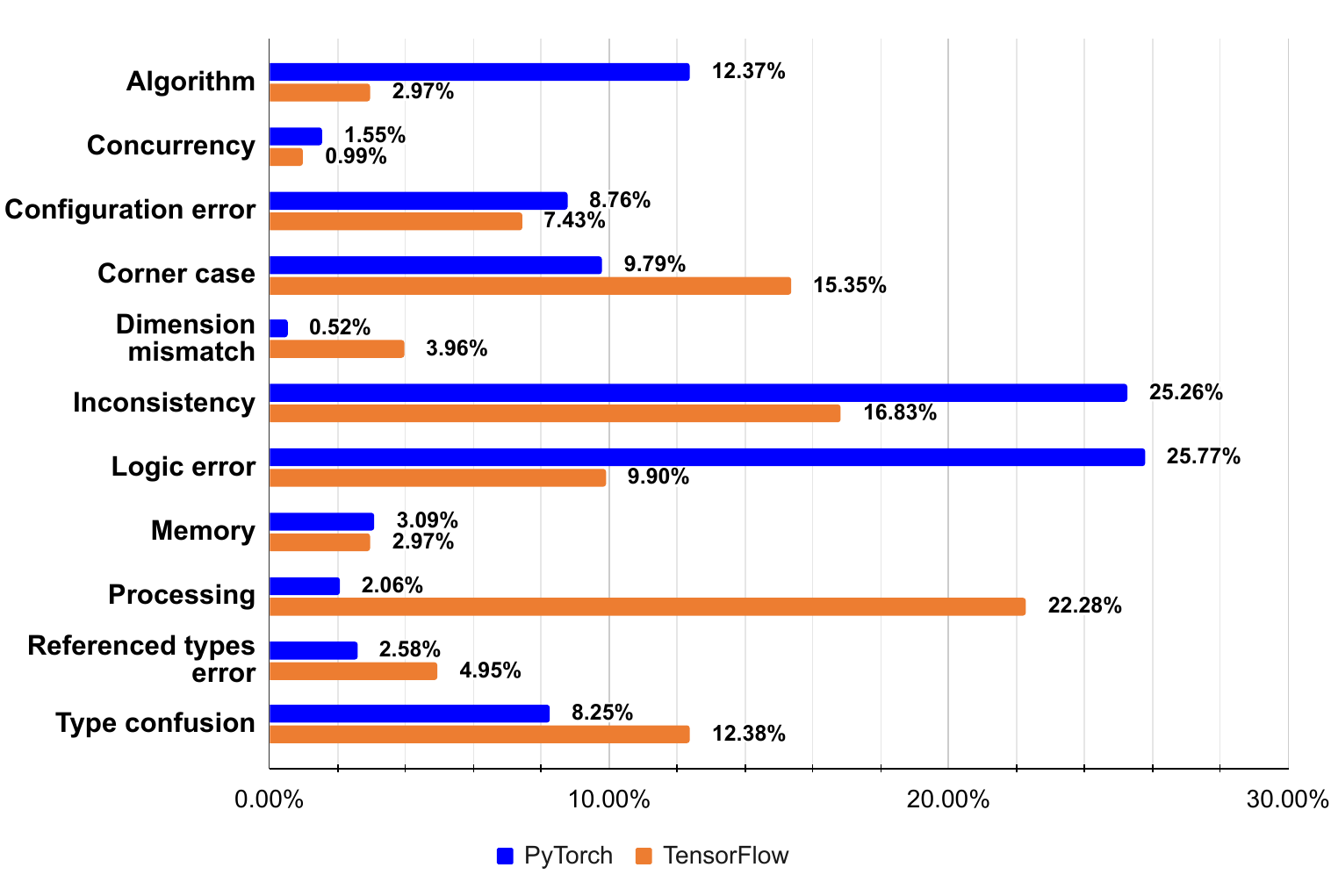}
        \vspace{-.5cm}
        \caption{Comparison of bug root causes between PyTorch and TensorFlow.}
        \label{fig:root-cause}
    \end{figure}
    \noindent
    
    \textbf{Comparison of bug root causes between TensorFlow and PyTorch.}
    Figure \ref{fig:root-cause} shows the comparison between our results and the results reported by Jia et al.~\cite{jia2021symptoms}.
    We first note a few similarities on the common bug root causes across both libraries. 
    Inconsistencies are the second most important bug root cause in both libraries, where changes in the APIs caused breaking changes or incompatible behaviour in the library. 
    Another common theme is the prevalence of type confusion bugs across both libraries, a common issue in dynamically typed languages such as Python and configuration errors, due to the highly configurable nature of both libraries~\cite{paszke2019pytorch,abadi2016tensorflow}. 
    There is, however, some stark contrast between frequencies of bug root causes. 
    Logic error (25\%) was the primary cause of bugs in PyTorch, while TensorFlow bugs were primarily caused by Processing-related issues (22\%). 
    Another major difference is that we find a much higher occurrence of bugs caused by wrong implementation of algorithms (12\% in PyTorch) than the figures reported in TensorFlow (3\%).

    \begin{tcolorbox}
        \textbf{Root Causes:} 
        PyTorch bugs are caused majorly by logic errors (25\%), API inconsistency (25\%), and wrong algorithm implementation (12\%). 
        Both PyTorch and TensorFlow share problems related to library API inconsistency and type-related issues, but TensorFlow exhibited a higher frequency of bugs caused by wrong variable initialization. 
    \end{tcolorbox}
    
    \subsection{Results: Symptoms of PyTorch Bugs}
    We have categorized the symptoms of bugs into 6 categories by following Jia et al. methodology~\cite{jia2021symptoms}.
    Table~\ref{tab:symptoms} shows that more than a third of bugs result in crashes during program execution. This is problematic since crashes cannot be detected during compile time and require the program to be executed, which can take a large amount of time. In the following, we discuss the symptoms in detail.

    \begin{table}
        \centering
        \caption{Symptoms of bugs identified in PyTorch.} 
        \label{tab:symptoms}
        
\begin{tabularx}{\linewidth}{l X r}
\toprule
   \textbf{Symptom} & \textbf{Description} & \textbf{Freq.} \\
\midrule
    Crash  & Irregular program exit & 35.05\%  \\
    Functional Error & Program doesn't function properly & 31.96\% \\
    Performance degradation & Poor performance or too much resource usage & 12.89\% \\
    Build Failure & Program fails to compile & 11.34\% \\
    Warning-style error & Display of warning message & 8.25\% \\
    Hang & Program gets stuck mid-run & 0.53\% \\
    
\bottomrule
\end{tabularx}

    \end{table}
    
    \begin{enumerate}
        \item\textbf{Crash (35.05\%).} The symptoms in this category involved program crashes. This means the program exits irregularly mid-execution by throwing an error. For example, issue \#973 \cite{issue973} reports a bug where the \texttt{DataLoader} causes a \texttt{RuntimeError} which causes the program to crash.
        
         \item\textbf{Functional error (31.96\%).} This symptom involves cases where the program fails to function properly as it was designed. For example, issue \#62967 \cite{issue62967} reports an issue caused by a bug which resulted in \texttt{BatchNorm} and \texttt{SyncBatchNorm} operations behaving incorrectly and causing the program to generate incorrect results.
         
        \item\textbf{Performance degradation (12.89\%).} This symptom involves cases where the program has poor performance or excessively uses compute resources e.g., takes an unreasonable amount of time to produce results. For example, issue \#25010 \cite{issue25010} reports such an instance where the CPU was under unreasonable load.
        
        \item\textbf{Build failure (11.34\%).} This symptom involves cases where the program fails to compile. There is a difference between crash and build failure symptoms. If a bug results in the program exiting mid-execution, it is considered a crash. However, if the program fails to compile, it is considered a build failure. 
        For example, in issue \#24175 \cite{issue24175}, it was reported that the program failed to compile because of a failed import caused by dependency issues.
        
        \item\textbf{Warning-style errors (8.25\%).} This symptom involves cases where the running program's execution was not interrupted. However, as reported in issue \#47038 \cite{issue47038}, the framework displayed error messages about feature deprecation or bad coding styles.
    \end{enumerate}
    
    The remaining symptom was not prevalent enough to be a cause for concern. The Hang symptom (0.52\%) such as issue \#48666 \cite{issue48666} were cases in which the program got stuck in the middle of a run without terminating or any response.\\
    
\begin{figure}
      \centering
      \includegraphics[width=\linewidth]{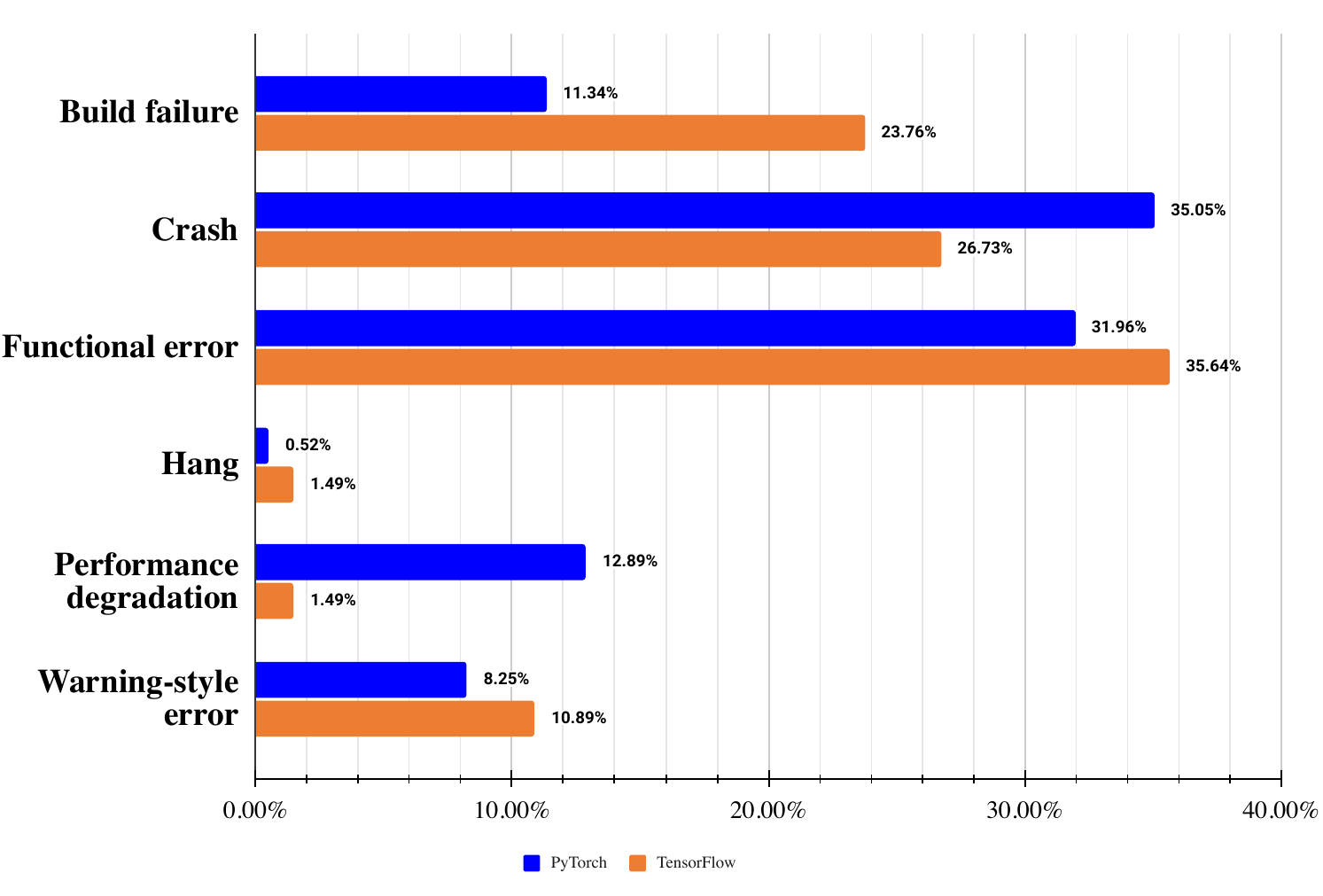}
        \vspace{-.5cm}
      \caption{Comparison of symptoms between PyTorch and TensorFlow.}
      \label{fig:symptoms}
    \end{figure}

    \noindent
    \textbf{Comparison of bug symptoms between TensorFlow and PyTorch.}
    We compare the frequency of bug symptoms in Figure \ref{fig:symptoms} across both PyTorch and TensorFlow. 
    Both libraries have, as the most frequent symptoms, bugs that cause functional errors and program crashes. 
    Build failure is significantly more common in the TensorFlow library (23\%), while PyTorch reports more frequent performance degradation (~13\%). 
    Warning-style errors are comparably similar, and bugs that cause the library to become not responsive are rare.

    \begin{tcolorbox}
        \textbf{Symptoms:} 
        Both PyTorch and TensorFlow frequently report as functional errors and program crash as the most frequent bug symptoms. 
        While PyTorch reports more frequent performance Degradation, build failures are reported more often in TensorFlow bugs. 
    \end{tcolorbox}

\section{Bug Locations}\label{rq2}

    \subsection{Motivation}
    DL libraries have many components and sub-systems such as different modules for data collection, data management, data validation, distributing training of models, etc. Analyzing where bugs have presented themselves the most, will assist the developers of PyTorch in improving the library's quality and pinpointing critical areas in need of extra attention. In addition, PyTorch users can better understand which components in the library may be more vulnerable to buggy behavior when using the API. Researchers may also be able to make design decisions for potential detector tools for DL libraries and software by knowing the target locations of the common bugs.
    
    \subsection{Methodology}
    To find bug locations, we first looked into issue descriptions and if the location was not specified, we looked into the commits associated with the issue to find its location. These locations may correspond to various functionalities and components in PyTorch. These functionalities and components were identified by looking at the official documentation, tutorials, and discussions in PyTorch's repository. However, we have disregarded superficial bugs, such as those related to poor code documentation.
    PyTorch does not formally declare its components in detail as an open-source project, but like other projects, it organizes its source files into distinct directories according to their functionalities. 
    
    \subsection{Results}

    \begin{table}
        \centering
        \caption{Location of bugs in PyTorch.} 
        \label{tab:components}
        \begin{tabularx}{\linewidth}{l X r}
\toprule
   \textbf{Component} & \textbf{Description} & \textbf{Freq.} \\
\midrule
    Core  & Defining operations for other modules & 40.59\%  \\
    Test & Testing framework's functionalities & 12.87\% \\
    Tools & Expanding framework functionalities, not developed by PyTorch team & 11.88\% \\
    Config & Configurations for framework's execution & 7.43\% \\
    Computation Graph & Computing tensor graph operations & 6.93\% \\
    CUDA & Interface with NVIDIA's CUDA & 6.93\% \\
    Documentation & Functionalities for describing other components & 4.95\% \\
    Framework & Functionalities that don't belong to other categories & 4.95\% \\
    API & Expand functionalities but not integrated into framework & 1.98\% \\
    NN Modules & Operations explicitly for NN functionality & 1.49\% \\
    
\bottomrule
\end{tabularx}

    \end{table}
    
    The following are the components that we have identified in PyTorch and the distribution of components affected by bugs is depicted in Table \ref{tab:components}.

    \begin{itemize}
      \item\textbf{Core (40.59\%)}. A component belongs to the core category if it is responsible for defining and handling operations upon which other modules and functionalities are built. For example, the ATen library is ``The foundational tensor and mathematical operation library on which all else is built."\cite{paszke2019pytorch} belongs to this category. It must be noted that the core component of PyTorch contains a lot of functionalities and therefore has a higher density than the other components.
      
      \item\textbf{Test (12.87\%)}. %
      PyTorch categorizes all scripts and functionalities that run a test or check other components in the \texttt{test} directory. Following their example, we categorize any bug that exists in any of the components in the \texttt{test} directory as bugs belonging to the test component. For example, PR \#34563 \cite{pr34563} modifies a function belonging to a script in this directory.
      
      \item\textbf{Tools (11.88\%)}. A component belongs to the tools category if it is used to expand the functionality of the framework but is not developed by PyTorch's development team. For example, Caffe \cite{jia2014caffe} which is integrated with PyTorch similar to how Keras is now integrated with TensorFlow2, belongs to this category.
      
      \item\textbf{Config (7.43\%)}. A component belongs to the config category if it is responsible for outlining configurations that define how the framework should execute. For example, scripts that determine dependency versions for each version of PyTorch, belong to this category.
      
      \item\textbf{Computation Graph (6.93\%)}. A component belongs to the computation graph category if it is responsible for computing tensor graphs and backpropagation operations. For example, ``AutoGrad" and ``optim" modules which are responsible for a part of the backpropagation process, belong to this category.

      \item\textbf{CUDA (6.93\%)}. The CUDA toolkit is provided by NVIDIA to allow for developing GPU-accelerated applications \cite{nvidia_cuda}. PyTorch uses the CUDA toolkit to allow developers to use GPUs instead of CPUs to accelerate the models' training process. A component belongs to the CUDA category if it is designed to work with the CUDA toolkit.

      \item\textbf{Documentation (4.95\%)}. 
      A component belongs to the documentation category if it is related to describing a component's functionality independent of the original component. Even though we do not consider bugs related to \texttt{code} documentation, a component might provide functionalities for understanding or describing how other components work. For example, PR \#43434 \cite{pr43434} fixes a bug in the definition of a datatype (\texttt{torch.nonzero}). Before fixing this documentation, the developers were not aware of this datatype's correct usage which resulted in the generation of redundant warnings.
      
      \item\textbf{Framework (4.95\%)}. A component belongs to the framework category if it is responsible for handling operations that are a part of the framework's functionality but do not belong to the core, tools, computation graph or Neural Network modules components. For example, the ``TensorBoard" modules belong to this category.

    \end{itemize}

    Two other components exhibited fewer occurrences of bugs. Bugs related to the API components occurred only 1.98\% of the time, and 1.49\% of the bugs were identified in the Neural Network component. \\

    \noindent
    \textbf{Comparison of Bug Components Between TensorFlow and PyTorch.}
    When programmers resolve issues, they frequently change different parts of the library, which are directly reflected in the core areas. %
    Compared to TensorFlow, PyTorch follows a different development structure and this difference is evident in the distribution of bug locations. Comparing with the results of Jia et al.\cite{jia2020empirical}, we can see that in TensorFlow most bugs belong to the \texttt{contribution} component which contains newly developed but not fully integrated features for TensorFlow; while in PyTorch the majority of bugs belong to the \texttt{core} category that contains PyTorch's underlying functionalities. While TensorFlow's second-most bug-prone component is the API component, the same for PyTorch is the \texttt{test} component. 

    \begin{tcolorbox}
        \textbf{Bug Locations:} 
        Bugs in PyTorch are reported across 10 major components. 
        Most bugs in PyTorch are reported in the core component of the library, where most functionalities are built upon.
        PyTorch's developers also frequently report bugs related to the test component, and in tool components developed by the community. 
    \end{tcolorbox}

\section{Repair Patterns}\label{rq3}

    \subsection{Motivation}
    Analyzing repair patterns for bugs inside PyTorch allows us to see if repair patterns and templates in PyTorch are comparable to those found by Jia et al. \cite{jia2021symptoms} for TensorFlow.
    Recurring repair patterns can be used to understand fix strategies, and help devise tools that harness such patterns to fix bugs in PyTorch in a semi-automated manner. 
    
    \subsection{Methodology}
    We identified the repair patterns by first inspecting the symptoms and root causes, and the files modified in the fixes of each bug. We then looked at files changes, line changes, and individual commits done to fix the bug. Afterwards, we analyzed the characteristics of the changes and potentially assign templates to bug fixes with similar repair patterns. These repair patterns were identified using the coding scheme based on the repair patterns identified by Jia et al. \cite{jia2021symptoms} for TensorFlow. These repair pattern categories were based on the prior studies of Kim et al. \cite{kim2013automatic}, Le et al. \cite{le2016history}, and Liu and Zhong \cite{liu2018mining} on repair patterns.

    We see that some issue fixes are repetitive, i.e., they appeared at least twice, thus we take the following steps to investigate them:
    \begin{enumerate}
      \item \textbf{Examining the symptoms and underlying reasons.} To define the general state of a bug, we examined the symptom, root cause, and location information gathered in previous sections.
      
      \item \textbf{Tracking down related code changes.} The number of connected files, altered lines, and commit frequency are all used to estimate the fix scale of a bug. If a commit contains changes that aren't related to fixing bugs (for example, test case changes), we ignored them. If a pull request addresses multiple bugs, we treated them as separate issues and examined them separately, however this is unusual in our experience. In general, individual templates are easier to find in small-scale fixes.
      
      \item \textbf{Examining the adjustments' features.} To describe the repair process in detail, we focus on several characteristics of a bug fix, including the scope of buggy code (in a method, a function Object() { [native code] }, or global), modified code elements (variables, methods, or classes), and modification intention (e.g., changing a value, and modifying if-statement conditions).
      
      \item \textbf{Fixing templates are extracted.} We believe that bug solutions with the same characteristics as those listed above could share the same repair template. We used these traits to construct repair patterns, and we used instances that appear several times as templates.
    \end{enumerate}

    \subsection{Results}
    It is important to note that we have omitted bugs in our dataset where the bug fix does not belong to an identifiable repair pattern (i.e., Isolated) and cases where the bug fixes are solely changes in how PyTorch is configured (i.e., Configuration). Out of the 194 bugs reviewed, 85 bugs were considered Isolated and 25 were considered Configuration. This accounts for about 57\% of all bugs analyzed. Because a large portion of the bugs had complex changes, it shows that many bugs in PyTorch involve multiple fixes that cannot be narrowed down to a specific template or pattern. A descriptive statistic of the identified repair patterns of the remaining 84 bugs is shown in Table \ref{tab:repairs} and we discuss the most recurring patterns in detail below:

    \begin{table}[t]
        \centering
        \caption{Repair patterns of bugs identified in PyTorch.} 
        \label{tab:repairs}
        \begin{tabularx}{\linewidth}{l X r}
\toprule
\textbf{Repair Pattern}     & \textbf{Description}                             & \textbf{Freq.} \\ \midrule
Value checker            & Adds a checking of a variable                    & 21.43\%          \\
Parameter modifier       & Changes in parameter of a function call          & 17.86\%          \\
Condition replacer       & Changes the condition in a conditional statement & 10.71\%          \\
Initializer modifier     & Changes how a variable in initialized            & 9.52\%          \\
Method replacer          & Replaces a method with another                   & 8.33\%           \\
Exception adder          & Adds an exception handling                       & 8.33\%           \\
Type replacer            & Replaces a type with another                     & 8.33\%           \\
Referenced type modifier & Changes an import statement                      & 5.95\%           \\
Variable replacer        & Replaces usage of a variable with another one    & 4.76\%           \\
Format checker           & Checks a data format                             & 2.38\%           \\
Exception modifier       & Changes an existing exception handling           & 2.38\%           \\ \bottomrule
\end{tabularx}

    \end{table}

        \begin{figure*}
      \centering
      \includegraphics[width=.7\linewidth]{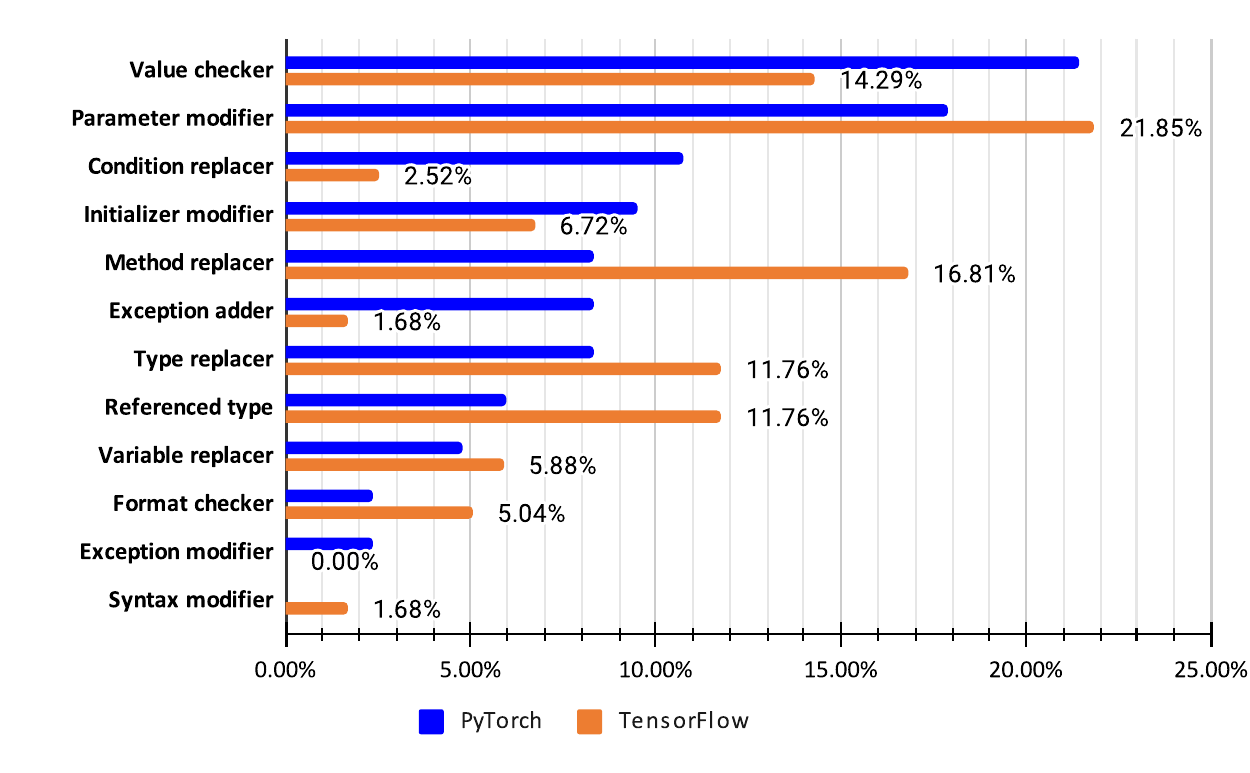}
      \vspace{-.5cm}
      \caption{Comparison of repair patterns between PyTorch and TensorFlow.}
      \label{fig:repair}
    \end{figure*}

    \begin{enumerate}
        \item \textbf{Value checker (21.43\%).}
        \normalsize This repair pattern involves checking a value of a variable. For example, issue \#61656 reports a crash when calculating the median of an empty tensor. Pull request \#61698 fixes this issue by returning a \texttt{NaN} by checking if the value of the \texttt{size} variable (representing the number of elements in the tensor) is less than or equal to zero.

        \item \textbf{Parameter modifier (17.86\%).}
            \normalsize This repair pattern involves modifying, replacing, adding or removing a parameter input. For example, PR \#38741 (see Listing~\ref{38741}) has the \texttt{equation} parameter passed to the function call to fix an incorrect operation. \\\\

\begin{lstlisting}[language=Python,basicstyle=\footnotesize,caption={Solution for the bug 38741.},captionpos=b,label=38741]
- return handle_torch_function(einsum, operands, *operands)
+ return handle_torch_function(einsum, operands, equation, *operands)
\end{lstlisting}

        \item \textbf{Condition replacer (10.71\%).}
            \normalsize This repair pattern involves replacing a predicate of a branch.
            For example, PR \#20759 adds an additional condition of \texttt{self.scalar\_type == src.scalar\_type()} when checking whether a tensor can properly do a copy and transpose, since type conversion from \texttt{float64} to \texttt{float32}  sometimes caused a crash. This was reported in issue \#20755 due to a bug introduced during code refactoring that was merged in PR \#20685.
            
        \item \textbf{Initializer modifier (9.52\%).}
            \normalsize This repair pattern involves modifying the initial value of a variable. For example, PR \#25111 initializes \texttt{pin\_memory thread} to only use 1 thread to fix issue \#25010 that describes high CPU utilization with \texttt{pin\_memory = True} and \texttt{num\_workers > 0} when using \texttt{DataLoader}.
            
        \item \textbf{Method replacer (8.33\%).}
            \normalsize This repair pattern involves replacing a method with another method with compatible parameters and return types.
            For example, PR \#22304 addresses \#21935 by using integer floor division for pooling shape computation that was introduced for convolution shapes in PR \#9640. This is done using the newly defined method \texttt{pooling\_output\_shape\_pad\_lr} that is called as part of \texttt{pooling\_output\_shape} that now acts as a wrapper that passes its parameter values to the aforementioned method, without changing the initial API.
            
        \item \textbf{Exception adder (8.33\%).}
            \normalsize This repair pattern adds handling or raising exceptions. For example, issue \#59312 reports that the \texttt{torch.utils.data.DataLoader} does not stop calling the \texttt{next} method despite the \texttt{IterableDataset} object having already raised a \texttt{StopIteration} exception. To resolve this, PR \#59313 raises the \texttt{StopIteration} exception again before the execution continues.
        
        \item \textbf{Type replacer (8.33\%).}
            \normalsize This repair pattern involves replacing the variable type. For example, the PR \#52828 addresses issue \#52822 that involves an illegal memory access when doing CUDA max pooling for large inputs by changing the type of the \texttt{slice} variable from \texttt{int} to \texttt{int64\_t} so that multiplication can be done in \texttt{int64\_t}.
        
        \item \textbf{Referenced type modifier (5.95\%).}
            \normalsize This repair pattern involves modifying, removing, replacing or adding a reference type. For example, the PR \#53424 addresses issue \#24807 by replacing the import from \texttt{tensor} to \texttt{\_tensor}  because the module \texttt{torch.tensor} is now \texttt{torch.\_tensor}.\\

    \end{enumerate}

    \noindent
    \textbf{Comparison of bug repair patterns between TensorFlow and PyTorch.}
    Figure~\ref{fig:repair} shows the comparison of recurring repair patterns across PyTorch and the patterns reported by Jia et al.~\cite{jia2021symptoms}.
    The most recurring repair patterns are common across both libraries: Value checker and Parameter modifier make the top of the more frequent fix strategies. 
    Some significant differences, however, emerge. 
    Condition replacer (10\% in PyTorch) appears significantly more frequent in PyTorch than in TensorFlow (2.5\%), while Method replacer appears as a frequent fix pattern for TensorFlow (16\%).
    An important overall difference, however, is that we only report finding recurrent patterns in a minority of bug fixes (84 out of 194), while Jia et al.~\cite{jia2021symptoms} did not report the number of bugs in which fix patterns were not easily identified.

    \begin{tcolorbox}
        \textbf{Repair Patterns:} 
        We find recurring patterns in 84 out of 194 bugs in PyTorch. 
        Our results show that Value checker (21.43\%), Parameter modifier (17.86\%), Condition replacer (10.71\%), and Initializer modifier (9.52\%) were commonly used repair patterns for bug fixing in PyTorch.
    \end{tcolorbox}

\section{Implications}\label{discussion}

\noindent
\textbf{Bugs in DL libraries have more in common with traditional software than with machine learning problems.}
While some bugs we analyze were caused by dimension mismatch, or wrong algorithm implementation, the vast majority of bugs were caused by Logic Errors and Inconsistent API changes (RQ1), problems that have long plagued library projects.
This suggests that, despite the unique nature of DL libraries, their maintenance faces some of the same challenges researchers have investigated in traditional software libraries. 
Approaches that target at identifying, e.g., breaking changes~\cite{breaking_changes} or better library update policies~\cite{abbas}, could greatly improve the reliability and maintainability of deep learning libraries and their applications.

\noindent
\textbf{Type confusion is a common issue in both PyTorch and TensorFlow libraries}.
This challenge can be largely attributed to Python's dynamic typing. 
While dynamic typing allows for more concise expressions in code, it also means that type-related bugs are often only discovered during runtime. 
The majority of bug symptoms we observed in these libraries were program crashes and functional errors, which can be disruptive and time-consuming to resolve. 
This highlights the need for more robust type checking mechanisms and better developer education on how to avoid type-related pitfalls in deep learning libraries~\cite{type_python}. 

\noindent
\textbf{Project organization influences bug location.}
The distinct project organization structures of PyTorch and TensorFlow play a significant role in the differences observed in bug locality across the two libraries. In the case of PyTorch, the majority of bugs were reported in the core component, which, as the name implies, is the most critical part of the framework. 
TensorFlow, on the other hand, allocates a separate component specifically for new features and external contributions. Consequently, this component attracts the majority of reported bugs.

\noindent
\textbf{Recurring fix patterns can be useful to devise automated strategies.}
Although we found recurring repair patterns in our analysis, their prevalence varied between the two libraries. In PyTorch, these patterns were responsible for fixing less than half of the analyzed bugs. Nevertheless, automated program repair approaches~\cite{le2016history} could potentially exploit the most common patterns to develop strategies that expedite the bug-fixing process in deep learning libraries.

\section{Threats to validity}\label{threats_to_validity}

\noindent
\textbf{Internal Validity.}
Eventual errors in manual labelling for each RQ could be considered an internal validity threat of our study. 
We mitigate this threat by relying on a well-defined code scheme from the replicated study, and employing two annotators on each bug and discussing eventual disagreements across all authors in the paper. 
Hence, we expect that our results hold even in the face of the inherent subjectivity of bug description interpretation.

\noindent
\textbf{External Validity.}
As discussed in Section \ref{methodology_section}, after our filtering, we had 2205 of bugs (including feature requests, documentation type issues, and super bugs) as candidates for analysis. Out of these candidates, we analyzed 194 bugs, which is comparable to the 200 bugs analyzed in the replicated study~\cite{jia2021symptoms}. %
Our findings may not mirror the distribution of symptoms, root causes and repair patterns of the entire PyTorch issues, but our filtering selection is meant to analyze the most impactful bugs of the library.

\section{Conclusion and future work}\label{conclusion}
In this paper, we conducted an empirical study of PyTorch bugs, which is one of the state-of-the-art DL libraries. We analyzed 194 bugs inside PyTorch and compared our findings with previous work that examined TensorFlow bugs. We found some similar patterns regarding the bug's root causes, symptoms, and repair patterns. However, our analysis also revealed a significant difference in terms of which components are prone to bugs. In the future, we plan to expand this study to cover more bugs. %
We also plan to examine the evolution of bugs patterns in %
DL libraries. %

\balance
\bibliography{refs}

\begin{thebibliography}{10}
\providecommand{\url}[1]{#1}
\csname url@samestyle\endcsname
\providecommand{\newblock}{\relax}
\providecommand{\bibinfo}[2]{#2}
\providecommand{\BIBentrySTDinterwordspacing}{\spaceskip=0pt\relax}
\providecommand{\BIBentryALTinterwordstretchfactor}{4}
\providecommand{\BIBentryALTinterwordspacing}{\spaceskip=\fontdimen2\font plus
\BIBentryALTinterwordstretchfactor\fontdimen3\font minus
  \fontdimen4\font\relax}
\providecommand{\BIBforeignlanguage}[2]{{%
\expandafter\ifx\csname l@#1\endcsname\relax
\typeout{** WARNING: IEEEtran.bst: No hyphenation pattern has been}%
\typeout{** loaded for the language `#1'. Using the pattern for}%
\typeout{** the default language instead.}%
\else
\language=\csname l@#1\endcsname
\fi
#2}}
\providecommand{\BIBdecl}{\relax}
\BIBdecl

\bibitem{Wang:HealthSystems}
\BIBentryALTinterwordspacing
H.~Wang, E.~Pujos-Guillot, B.~Comte, J.~L. de~Miranda, V.~Spiwok, I.~Chorbev,
  F.~Castiglione, P.~Tieri, S.~Watterson, R.~McAllister, T.~de~Melo~Malaquias,
  M.~Zanin, T.~S. Rai, and H.~Zheng, ``{Deep learning in systems medicine},''
  \emph{Briefings in Bioinformatics}, vol.~22, no.~2, pp. 1543--1559, 11 2020.
  [Online]. Available: \url{https://doi.org/10.1093/bib/bbaa237}
\BIBentrySTDinterwordspacing

\bibitem{bojarski2016end}
M.~Bojarski, D.~Del~Testa, D.~Dworakowski, B.~Firner, B.~Flepp, P.~Goyal, L.~D.
  Jackel, M.~Monfort, U.~Muller, J.~Zhang \emph{et~al.}, ``End to end learning
  for self-driving cars,'' \emph{arXiv preprint arXiv:1604.07316}, 2016.

\bibitem{paszke2019pytorch}
A.~Paszke, S.~Gross, F.~Massa, A.~Lerer, J.~Bradbury, G.~Chanan, T.~Killeen,
  Z.~Lin, N.~Gimelshein, L.~Antiga \emph{et~al.}, ``Pytorch: An imperative
  style, high-performance deep learning library,'' \emph{Advances in neural
  information processing systems}, vol.~32, 2019.

\bibitem{abadi2016tensorflow}
M.~Abadi, P.~Barham, J.~Chen, Z.~Chen, A.~Davis, J.~Dean, M.~Devin,
  S.~Ghemawat, G.~Irving, M.~Isard \emph{et~al.}, ``$\{$TensorFlow$\}$: A
  system for $\{$Large-Scale$\}$ machine learning,'' in \emph{12th USENIX
  symposium on operating systems design and implementation (OSDI 16)}, 2016,
  pp. 265--283.

\bibitem{StackOve55:online}
S.~Overflow, ``Developer survey 2022,''
  \url{https://survey.stackoverflow.co/2022/#most-popular-technologies-misc-tech-prof},
  May 2022, (Accessed on 04/02/2023).

\bibitem{zhang2018empirical}
Y.~Zhang, Y.~Chen, S.-C. Cheung, Y.~Xiong, and L.~Zhang, ``An empirical study
  on tensorflow program bugs,'' in \emph{Proceedings of the 27th ACM SIGSOFT
  International Symposium on Software Testing and Analysis}, 2018, pp.
  129--140.

\bibitem{islam2019comprehensive}
M.~J. Islam, G.~Nguyen, R.~Pan, and H.~Rajan, ``A comprehensive study on deep
  learning bug characteristics,'' in \emph{Proceedings of the 2019 27th ACM
  Joint Meeting on European Software Engineering Conference and Symposium on
  the Foundations of Software Engineering}, 2019, pp. 510--520.

\bibitem{humbatova2020taxonomy}
N.~Humbatova, G.~Jahangirova, G.~Bavota, V.~Riccio, A.~Stocco, and P.~Tonella,
  ``Taxonomy of real faults in deep learning systems,'' in \emph{Proceedings of
  the ACM/IEEE 42nd International Conference on Software Engineering}, 2020,
  pp. 1110--1121.

\bibitem{tambon2021silent}
F.~Tambon, A.~Nikanjam, L.~An, F.~Khomh, and G.~Antoniol, ``Silent bugs in deep
  learning frameworks: An empirical study of keras and tensorflow,''
  \emph{arXiv preprint arXiv:2112.13314}, 2021.

\bibitem{jia2021symptoms}
L.~Jia, H.~Zhong, X.~Wang, L.~Huang, and X.~Lu, ``The symptoms, causes, and
  repairs of bugs inside a deep learning library,'' \emph{Journal of Systems
  and Software}, vol. 177, p. 110935, 2021.

\bibitem{jia2020empirical}
------, ``An empirical study on bugs inside tensorflow,'' in
  \emph{International Conference on Database Systems for Advanced
  Applications}.\hskip 1em plus 0.5em minus 0.4em\relax Springer, 2020, pp.
  604--620.

\bibitem{pytorchtesla}
\BIBentryALTinterwordspacing
A.~Karpathy, ``Pytorch at tesla - andrej karpathy, tesla.'' [Online].
  Available: \url{https://youtu.be/oBklltKXtDE}
\BIBentrySTDinterwordspacing

\bibitem{goodman2019uber}
\BIBentryALTinterwordspacing
N.~Goodman, ``Uber ai labs open sources pyro, a deep probabilistic programming
  language,'' Feb 2019. [Online]. Available: \url{https://eng.uber.com/pyro/}
\BIBentrySTDinterwordspacing

\bibitem{pradel2018deepbugs}
M.~Pradel and K.~Sen, ``Deepbugs: A learning approach to name-based bug
  detection,'' \emph{Proceedings of the ACM on Programming Languages}, vol.~2,
  no. OOPSLA, pp. 1--25, 2018.

\bibitem{lu2005bugbench}
S.~Lu, Z.~Li, F.~Qin, L.~Tan, P.~Zhou, and Y.~Zhou, ``Bugbench: Benchmarks for
  evaluating bug detection tools,'' in \emph{Workshop on the evaluation of
  software defect detection tools}, vol.~5.\hskip 1em plus 0.5em minus
  0.4em\relax Chicago, Illinois, 2005.

\bibitem{wang2016bugram}
S.~Wang, D.~Chollak, D.~Movshovitz-Attias, and L.~Tan, ``Bugram: bug detection
  with n-gram language models,'' in \emph{Proceedings of the 31st IEEE/ACM
  International Conference on Automated Software Engineering}, 2016, pp.
  708--719.

\bibitem{pan2006bug}
K.~Pan, S.~Kim, and E.~J. Whitehead~Jr, ``Bug classification using program
  slicing metrics,'' in \emph{2006 Sixth IEEE International Workshop on Source
  Code Analysis and Manipulation}.\hskip 1em plus 0.5em minus 0.4em\relax IEEE,
  2006, pp. 31--42.

\bibitem{zhou2016combining}
Y.~Zhou, Y.~Tong, R.~Gu, and H.~Gall, ``Combining text mining and data mining
  for bug report classification,'' \emph{Journal of Software: Evolution and
  Process}, vol.~28, no.~3, pp. 150--176, 2016.

\bibitem{d2010extensive}
M.~D'Ambros, M.~Lanza, and R.~Robbes, ``An extensive comparison of bug
  prediction approaches,'' in \emph{2010 7th IEEE working conference on mining
  software repositories (MSR 2010)}.\hskip 1em plus 0.5em minus 0.4em\relax
  IEEE, 2010, pp. 31--41.

\bibitem{hammouri2018software}
A.~Hammouri, M.~Hammad, M.~Alnabhan, and F.~Alsarayrah, ``Software bug
  prediction using machine learning approach,'' \emph{International Journal of
  Advanced Computer Science and Applications}, vol.~9, no.~2, pp. 78--83, 2018.

\bibitem{arcuri2008novel}
A.~Arcuri and X.~Yao, ``A novel co-evolutionary approach to automatic software
  bug fixing,'' in \emph{2008 IEEE Congress on Evolutionary Computation (IEEE
  World Congress on Computational Intelligence)}.\hskip 1em plus 0.5em minus
  0.4em\relax IEEE, 2008, pp. 162--168.

\bibitem{ye2021comprehensive}
H.~Ye, M.~Martinez, T.~Durieux, and M.~Monperrus, ``A comprehensive study of
  automatic program repair on the quixbugs benchmark,'' \emph{Journal of
  Systems and Software}, vol. 171, p. 110825, 2021.

\bibitem{thung2012empirical}
F.~Thung, S.~Wang, D.~Lo, and L.~Jiang, ``An empirical study of bugs in machine
  learning systems,'' in \emph{2012 IEEE 23rd International Symposium on
  Software Reliability Engineering}.\hskip 1em plus 0.5em minus 0.4em\relax
  IEEE, 2012, pp. 271--280.

\bibitem{kim2021denchmark}
M.~Kim, Y.~Kim, and E.~Lee, ``Denchmark: A bug benchmark of deep
  learning-related software,'' in \emph{2021 IEEE/ACM 18th International
  Conference on Mining Software Repositories (MSR)}.\hskip 1em plus 0.5em minus
  0.4em\relax IEEE, 2021, pp. 540--544.

\bibitem{jia2014caffe}
Y.~Jia, E.~Shelhamer, J.~Donahue, S.~Karayev, J.~Long, R.~Girshick,
  S.~Guadarrama, and T.~Darrell, ``Caffe: Convolutional architecture for fast
  feature embedding,'' in \emph{Proceedings of the 22nd ACM international
  conference on Multimedia}, 2014, pp. 675--678.

\bibitem{keras}
K.~Team, ``Keras: Deep learning for humans,''
  \url{https://github.com/keras-team/keras}, 2015.

\bibitem{bergstra2011theano}
J.~Bergstra, F.~Bastien, O.~Breuleux, P.~Lamblin, R.~Pascanu, O.~Delalleau,
  G.~Desjardins, D.~Warde-Farley, I.~Goodfellow, A.~Bergeron \emph{et~al.},
  ``Theano: Deep learning on gpus with python,'' in \emph{NIPS 2011,
  BigLearning Workshop, Granada, Spain}, vol.~3.\hskip 1em plus 0.5em minus
  0.4em\relax Citeseer, 2011, pp. 1--48.

\bibitem{collobert2002torch}
R.~Collobert, S.~Bengio, and J.~Mari{\'e}thoz, ``Torch: a modular machine
  learning software library,'' Idiap, Tech. Rep., 2002.

\bibitem{zhang2020towards}
X.~Zhang, X.~Xie, L.~Ma, X.~Du, Q.~Hu, Y.~Liu, J.~Zhao, and M.~Sun, ``Towards
  characterizing adversarial defects of deep learning software from the lens of
  uncertainty,'' in \emph{2020 IEEE/ACM 42nd International Conference on
  Software Engineering (ICSE)}.\hskip 1em plus 0.5em minus 0.4em\relax IEEE,
  2020, pp. 739--751.

\bibitem{chen2022toward}
J.~Chen, Y.~Liang, Q.~Shen, J.~Jiang, and S.~Li, ``Toward understanding deep
  learning framework bugs,'' \emph{ACM Transactions on Software Engineering and
  Methodology}, 2022.

\bibitem{Andrew:CollaborativeCoding}
\BIBentryALTinterwordspacing
K.~A.~R. Richards and M.~A. Hemphill, ``A practical guide to collaborative
  qualitative data analysis,'' \emph{Journal of Teaching in Physical
  Education}, vol.~37, no.~2, pp. 225 -- 231. [Online]. Available:
  \url{https://journals.humankinetics.com/view/journals/jtpe/37/2/article-p225.xml}
\BIBentrySTDinterwordspacing

\bibitem{issue3307:copydeep21}
G.~Roeder, ``Issue \#3307: `copy.deepcopy` does not copy gradient buffers of
  `torch.autograd.variable` instance.''
  \url{https://github.com/pytorch/pytorch/issues/3307}, Oct 2017, (Accessed on
  09/20/2022).

\bibitem{pr53424}
B.~Hirsh, ``Pull request \#53424: `[pytorch][pr] fix pylint error torch.tensor
  is not callable `.'' \url{https://github.com/pytorch/pytorch/pull/53424},
  March 2021, (Accessed on 09/25/2022).

\bibitem{pr56488}
E.~Guan, ``Pull request \#56488: `[dataloader] add numpy seeding to worker of
  dataloader`.'' \url{https://github.com/pytorch/pytorch/pull/56488}, April
  2021, (Accessed on 09/25/2022).

\bibitem{issue16532}
Travis, ``Issue \#16532: `missing gradient when autograd called inside a
  function on multi-gpu (eg gradient penalty)`.''
  \url{https://github.com/pytorch/pytorch/issues/16532}, Jan 2019, (Accessed on
  09/25/2022).

\bibitem{issue22389}
V.~Kantorov, ``Issue \#22389: `dependency issues with torch.utils.tensorboard:
  "no module named past" and "no module named 'pil'" `.''
  \url{https://github.com/pytorch/pytorch/issues/22389}, Jul 2019, (Accessed on
  09/25/2022).

\bibitem{issue42218}
J.~Jeong, ``Issue \#42218: `type mismatch error with
  torch.nn.functional.grid\_sample under amp `.''
  \url{https://github.com/pytorch/pytorch/issues/42218}, Jul 2020, (Accessed on
  09/25/2022).

\bibitem{issue35901}
D.~MacLeod, ``Issue \#35901: `oom error where ~50\% of the gpu ram cannot be
  utilised/reserved`.'' \url{https://github.com/pytorch/pytorch/issues/35901},
  Apr 2020, (Accessed on 09/25/2022).

\bibitem{issue42435}
A.~Oke, ``Issue \#42435: `cannot import swa utils`.''
  \url{https://github.com/pytorch/pytorch/issues/42435}, Aug 2020, (Accessed on
  09/25/2022).

\bibitem{issue49052}
H.~Huang, ``Issue \#49052: `tcpstore constructor arguments mismatch unexpected
  behavior`.'' \url{https://github.com/pytorch/pytorch/issues/49052}, Dec 2020,
  (Accessed on 09/25/2022).

\bibitem{issue67626}
mrshenli, ``Issue \#67626: `get rid of the blocking call in rrefproxy`.''
  \url{https://github.com/pytorch/pytorch/issues/67626}, Nov 2021, (Accessed on
  09/25/2022).

\bibitem{issue71065}
H.~Liao, ``Issue \#71065: `fix the shape inconsistency of out and elem
  tensor`.'' \url{https://github.com/pytorch/pytorch/pull/71065}, Jan 2022,
  (Accessed on 09/25/2022).

\bibitem{issue973}
J.~F. Santos, ``Issue \#973: `file\_descriptor sharing strategy may be leaking
  fds, resulting in dataloader causing runtimeerror: received 0 items of
  ancdata`.'' \url{https://github.com/pytorch/pytorch/issues/973}, Mar 2017,
  (Accessed on 10/02/2022).

\bibitem{issue62967}
E.~Rees, ``Issue \#62967: Batchnorm2d + syncbatchnorm incorrect multi gpu
  behaviour in 1.10.0 (tested working in 1.8.0).''
  \url{https://github.com/pytorch/pytorch/issues/62967}, Aug 2021, (Accessed on
  10/02/2022).

\bibitem{issue25010}
R.~Wightman, ``Issue \#25010: Very high cpu utilization with pin\_memory=true
  and num\_workers \> 0.''
  \url{https://github.com/pytorch/pytorch/issues/25010}, Aug 2019, (Accessed on
  10/02/2022).

\bibitem{issue24175}
M.~Iqbal, ``Issue \#24175: Error caffe2.python when tried importing
  summarywriter from torch.utils.tensorboard.''
  \url{https://github.com/pytorch/pytorch/issues/24175}, Aug 2019, (Accessed on
  10/02/2022).

\bibitem{issue47038}
C.~Mocholí, ``Issue \#47038: Pytorch 1.7.0 cuda driver warning.''
  \url{https://github.com/pytorch/pytorch/issues/47038}, Oct 2020, (Accessed on
  10/02/2022).

\bibitem{issue48666}
T.~Wang, ``Issue \#48666: [dataloader] hang on python exit when has iter ref
  and sampler yields large indices.''
  \url{https://github.com/pytorch/pytorch/issues/48666}, Dec 2020, (Accessed on
  10/02/2022).

\bibitem{pr34563}
tugrulince, ``Pull request \#34563: `support left and right shift operators in
  jit`.'' \url{https://github.com/pytorch/pytorch/pull/34563}, April 2021,
  (Accessed on 04/25/2023).

\bibitem{nvidia_cuda}
NVIDIA, ``Nvidia cuda,'' \url{https://developer.nvidia.com/cuda-toolkit},
  (Accessed on 05/25/2023).

\bibitem{pr43434}
S.~Bekman, ``Pull request \#43434: `[doc] correct docs for torch.nonzero`.''
  \url{https://github.com/pytorch/pytorch/pull/43434}, April 2021, (Accessed on
  05/25/2023).

\bibitem{kim2013automatic}
D.~Kim, J.~Nam, J.~Song, and S.~Kim, ``Automatic patch generation learned from
  human-written patches,'' in \emph{2013 35th International Conference on
  Software Engineering (ICSE)}.\hskip 1em plus 0.5em minus 0.4em\relax IEEE,
  2013, pp. 802--811.

\bibitem{le2016history}
X.~B.~D. Le, D.~Lo, and C.~Le~Goues, ``History driven program repair,'' in
  \emph{2016 IEEE 23rd international conference on software analysis,
  evolution, and reengineering (SANER)}, vol.~1.\hskip 1em plus 0.5em minus
  0.4em\relax IEEE, 2016, pp. 213--224.

\bibitem{liu2018mining}
X.~Liu and H.~Zhong, ``Mining stackoverflow for program repair,'' in \emph{2018
  IEEE 25th international conference on software analysis, evolution and
  reengineering (SANER)}.\hskip 1em plus 0.5em minus 0.4em\relax IEEE, 2018,
  pp. 118--129.

\bibitem{breaking_changes}
A.~Brito, L.~Xavier, A.~Hora, and M.~T. Valente, ``Apidiff: Detecting api
  breaking changes,'' in \emph{2018 IEEE 25th International Conference on
  Software Analysis, Evolution and Reengineering (SANER)}, 2018, pp. 507--511.

\bibitem{abbas}
A.~J. Jafari, D.~E. Costa, R.~Abdalkareem, E.~Shihab, and N.~Tsantalis,
  ``Dependency smells in javascript projects,'' \emph{IEEE Transactions on
  Software Engineering}, vol.~48, no.~10, pp. 3790--3807, 2022.

\bibitem{type_python}
F.~Khan, B.~Chen, D.~Varro, and S.~McIntosh, ``An empirical study of
  type-related defects in python projects,'' \emph{IEEE Transactions on
  Software Engineering}, vol.~48, no.~8, pp. 3145--3158, 2022.

\end{thebibliography}

\end{document}